\documentstyle[12pt,aps]{revtex}
\begin{document}

\title{
\begin{flushright}
\vspace{-1cm}
{\normalsize MC/TH 96/12}
\vspace{1cm}
\end{flushright}
Chiral symmetry and QCD sum rules\\ for the nucleon mass in matter\bigskip}
\author{Michael C. Birse and Boris Krippa\thanks{Permanent address: Institute 
for Nuclear Research of the Russian Academy of Sciences, Moscow Region 117312,
Russia.}}
\address{Theoretical Physics Group, Department of Physics and Astronomy\\
University of Manchester, Manchester, M13 9PL, UK\\}
\maketitle
\bigskip
\bigskip

\begin{abstract}
The quark condensate in nuclear matter contains a term of order $\rho m_\pi$,
arising from the contribution of low-momentum virtual pions to the $\pi N$
sigma commutator. Standard treatments of QCD sum rules for a nucleon in matter
generate a similar term in the nucleon effective mass, although this is
inconsistent with chiral perturbation theory. We show how an improved
treatment of pionic contributions on the phenomenological side of the sum
rules can cancel out this unwanted piece. Our results also show that
factorisation ansatz for the four-quark condensate cannot be valid in matter.
\end{abstract}
\pacs{11.30.Rd, 11.55.Hx, 21.65.+f}

The presence of dense nuclear matter could significantly influence the QCD
vacuum, for example by partially restoring the chiral symmetry of QCD. There is
much interest at present in elucidating the effects this might have on the
properties of nucleons and other hadrons in matter\cite{csrevs}.
In this context, the method of QCD sum rules\cite{svz} provides a promising
approach to connecting hadron properties with underlying QCD matrix elements.
Applications of this to the energy of a nucleon in matter\cite{nmsr} have shown
that changes to the condensates in matter can give rise to large scalar and
vector self-energies that are similar to those in relativistic nuclear
phenomenology\cite{dirph}.

A major ingredient in QCD sum rules for the nucleon mass is the scalar quark
condensate. To first order in the density the change in this condensate in
matter is given by the model-independent result\cite{dl,cfg}
\begin{equation}
{\langle\Psi|\overline qq|\Psi\rangle\over\langle 0|\overline qq|0\rangle}
=1-{\sigma_{\pi{\scriptscriptstyle N}}\over f_\pi^2 m_\pi^2}\rho,
\label{ratio}
\end{equation}
where where $|\Psi\rangle$ denotes our nuclear matter ground-state and
$\sigma_{\pi{\scriptscriptstyle N}}$ is the pion-nucleon sigma commutator. The
leading nonanalytic dependence of the sigma commutator on the current quark
mass $\overline m$ is given by a term of order $\overline m^{3/2} \propto
m_\pi^3$\cite{gl}. This arises from the long-ranged part of the pion cloud
of the nucleon.

The leading correction term in the chiral expansion of the density-dependent
part of the quark condensate in matter is thus of order $\rho m_\pi$. However
chiral counting rules\cite{wein} tell us that no term of order $m_\pi$ appears
in the $NN$ interaction\cite{orvk} and hence that no term of this order should
be present in the self-energy of a nucleon in matter. As one of us has pointed
out recently\cite{birqq}, the nucleon mass in matter thus cannot simply be a
function of the quark condensate alone. Other long-ranged effects must
contribute to ensure that no term of order $\rho m_\pi$ appears. This presents
a potential problem for QCD sum rules for a nucleon in matter, since the
operator-product expansion (OPE) involves matrix elements of local operators
like $\overline qq$ which cannot distinguish between short- and long-ranged
contributions. The cancellation of the parts of these condensates that are of
order $\rho m_\pi$ must therefore come from the phenomenological side of the
sum rules, and in particular from states involving low-momentum pions. Similar
cancellations have been found previously in sum rules for the mass of a
nucleon at finite temperature\cite{el,koi} and in vacuum\cite{lccg}.

Sum rules are obtained by matching two expressions for a correlator of 
interpolating fields with the quantum numbers of the particle of interest. The
first of these is an OPE of the correlator at large (space-like) momentum which
introduces information about the nonperturbative features of the QCD ground
state (either the vacuum or matter) through the various condensates that
appear in it. The second takes the form a dispersion relation based on a
spectral representation of the correlator, for which one makes a simple
phenomenological model. This representation contains a pole corresponding to
the hadron state whose properties one is interested in, as well as a continuum
of higher-energy states with the same quantum numbers. Standard sum rule
analyses assume that the continuum below some threshold energy can be
neglected. They include only a high-energy continuum, which is modelled by the
QCD continuum from the OPE. Current applications of the method to a nucleon in
matter thus give rise to an effective mass containing a term of order $\rho
m_\pi$, produced by the corresponding term in the quark condensate. We show
here how, with a better treatment of the continuum, this term can be cancelled
by contributions of the same order from states with low-momentum pions.

The leading nonanalytic term in the sigma commutator, which we denote by 
$\sigma_{\pi{\scriptscriptstyle N}}^{(3)}$, has the form\cite{gl}
\begin{equation}
\sigma_{\pi{\scriptscriptstyle N}}^{(3)}=-{9\over 16\pi}\left({g_{\pi 
{\scriptscriptstyle NN}}\over 2 M_{\scriptscriptstyle N}}\right)^2m_\pi^3.
\end{equation}
It is produced by the lowest-momentum virtual pions in the cloud surrounding
the nucleon. Since these are very-low momentum pions, their interactions with a
second nucleon are suppressed by two powers of $m_\pi$ relative to their
contributions to the scalar quark density ($\sigma_{\pi{\scriptscriptstyle
N}}^{(3)}/m_\pi^2$) and so the leading nonanalytic terms in the $NN$
interaction are of order $m_\pi^3$\cite{wein,orvk}. At mean-field level, these
terms in the interaction give rise to leading nonanalytic terms in the
self-energy of a nucleon in matter of order $\rho m_\pi^3$ and not $\rho
m_\pi$.

One can relate $\sigma_{\pi{\scriptscriptstyle N}}^{(3)}$ to the leading
nonanalytic contribution of the pion mass term to the nucleon mass
(cf.\cite{bmcg})
\begin{equation}
{\textstyle{1\over 2}}m_\pi^2\langle N(\hbox{\bf p})|
:\hbox{\boldmath$\pi$}(0)^2:|N(\hbox{\bf p}')\rangle\simeq
(2\pi)^{-3}\sigma_{\pi{\scriptscriptstyle N}}^{(3)}+\cdots,
\label{signonan}
\end{equation}
where {\boldmath$\pi$}({\bf r}) is the pion field.\footnote{The nucleon states 
here are normalised so that $\langle N(\hbox{\bf p})|N(\hbox{\bf p}')\rangle
=(E_p/M)\delta(\hbox{\bf p}-\hbox{\bf p}')$. Only the leading term 
in powers of $m_\pi$ and momenta has been written out in (\ref{signonan}).}
There should, in principle, also be a term of order $m_\pi^2$ in this matrix
element. This has been ``renormalised away" by lumping it with other terms of
the same order, all of which contribute to the quark condensate at order
$m_\pi^0$. We can therefore to relate the ratio in (\ref{ratio}) to its value
in the chiral limit $\overline m=0$ by
\begin{equation}
{\langle\Psi|\overline qq|\Psi\rangle\over\langle 0|\overline qq|0\rangle}
\simeq\left.{\langle\Psi|\overline qq|\Psi\rangle\over\langle 0|\overline qq
|0\rangle}\right|_{\overline m=0}(1-\xi),
\label{qqlc}
\end{equation}
where we have defined
\begin{equation}
\xi={\rho\sigma_{\pi{\scriptscriptstyle N}}^{(3)}\over f_\pi^2 m_\pi^2}.
\end{equation}

Sum rules for a nucleon in matter are obtained from the correlator of two
interpolating nucleon fields\cite{iof}
\begin{equation}
\Pi(p)=i\!\int d^4x e^{ip\cdot x}\langle \Psi|T\{\eta(x)\,\overline{\eta}(0)\}
|\Psi\rangle.
\label{correl}
\end{equation}
Since we are interested in the leading chiral corrections to this, we need to
examine the piece of this correlator that is of order $\rho m_\pi$. These
corrections arise from pieces of the correlator where the interpolating fields
couple directly to a low-momentum pion from the cloud of one of the nucleons
forming the matter. A convenient way to obtain these terms is to use partial
conservation of the axial current (PCAC)\cite{dgh} in the way suggested by
Eletsky\cite{el} (see also\cite{dei,koi}). The main difference is that in the
present case we are dealing with virtual pions and so all possible
time-orderings need to be considered.

To the order we are interested, the piece of the correlator where the ``probe"
nucleon created by the interpolating fields interacts with a pion that is
emitted and then absorbed by the matter can be written in the form
\begin{equation}
\sum_{a,b}\int {d^3\hbox{\bf k}\over 2\omega_k}
{d^3\hbox{\bf k}'\over 2\omega_{k'}}\langle\Psi|a^{a\dagger}(\hbox{\bf k})
a^b(\hbox{\bf k}')|\Psi\rangle
\,i\!\int d^4x e^{ip\cdot x}\langle \Psi\,\pi^a(\hbox{\bf k})|T\{\eta(x)\,
\overline{\eta}(0)\}|\Psi\,\pi^b(\hbox{\bf k}')\rangle,
\label{piinout}
\end{equation}
similar to that needed in the case of thermal correction. In addition, we need
the other time-orderings where two pions are either emitted or absorbed by the
matter. These are, respectively,\footnote{In these expressions
we have used a covariant normalisation for the pions such that $\langle
\pi^a(\hbox{\bf k})|\pi^b(\hbox{\bf k}')\rangle=2\omega_k\delta^{ab}
\delta(\hbox{\bf k}-\hbox{\bf k}')$.}
\begin{equation}
{1\over 2}\sum_{a,b}\int {d^3\hbox{\bf k}\over 2\omega_k}
{d^3\hbox{\bf k}'\over 2\omega_{k'}}\langle\Psi|a^a(\hbox{\bf k})
a^b(\hbox{\bf k}')|\Psi\rangle
\,i\!\int d^4x e^{ip\cdot x}\langle \Psi|T\{\eta(x)\,\overline{\eta}(0)\}
|\Psi\,\pi^a(\hbox{\bf k})\pi^b(\hbox{\bf k}')\rangle,
\label{twopiin}
\end{equation}
and
\begin{equation}
{1\over 2}\sum_{a,b}\int {d^3\hbox{\bf k}\over 2\omega_k}
{d^3\hbox{\bf k}'\over 2\omega_{k'}}\langle\Psi|a^{a\dagger}(\hbox{\bf k})
a^{b\dagger}(\hbox{\bf k}')
|\Psi\rangle
\,i\!\int d^4x e^{ip\cdot x}\langle \Psi\,\pi^a(\hbox{\bf k})\pi^b(\hbox{\bf 
k}')|
T\{\eta(x)\,\overline{\eta}(0)\}|\Psi\rangle
\label{twopiout}
\end{equation}
where $\omega_k=\sqrt{\hbox{\bf k}^2+m_\pi^2}$.

Note we are ignoring pieces of the correlator where the pions transfer nonzero
momentum to the probe nucleon and that momentum is transfered back by another
interaction while the nucleon is propagating through the matter. Such terms
give rise to two-loop diagrams, with two integrals over pion momenta, and so
are of higher-order in the chiral expansion. This has allowed us to write
the matrix elements in the factorised forms 
(\ref{piinout},\ref{twopiin},\ref{twopiout}), with the ground state of nuclear 
matter $|\Psi\rangle$ appearing in in all of them.

For the leading chiral correction we need only the contributions of the
lowest-momentum pions in the cloud, and hence we may approximate the matrix
elements of the interpolating fields in the expressions above by their values
in the soft-pion limit\cite{el}. In this limit we can use the soft-pion
theorem of PCAC\cite{dgh} to rewrite (\ref{piinout}) in terms of
\begin{equation}
i\int d^4x e^{ip\cdot x}\langle \Psi\,\pi^a(\hbox{\bf k})|T\{\eta(x)\,
\overline{\eta}(0)\}|\Psi\,\pi^b(\hbox{\bf k}')\rangle
\simeq {-i\over f_\pi^2}\int d^4x e^{ip\cdot x}\langle \Psi|\left[Q_5^a,
\left[Q_5^b, T\{\eta(x)\,\overline{\eta}(0)\}\right]\right]|\Psi\rangle,
\label{soft}
\end{equation}
while the matrix elements in both (\ref{twopiin}) and (\ref{twopiout}) also
reduce to the same expression. If we use Ioffe's choice of interpolating field
for the nucleon\cite{iof}, this transforms under axial isospin rotations
as\cite{el}
\begin{equation}
\left[Q_5^a, \eta(x)\right]=-\gamma_5{\tau^a\over 2}\eta(x).
\label{fchtr}
\end{equation}
The matrix element of the double commutator in (\ref{soft}) can then be 
reduced to
\begin{equation}
i\int d^4x e^{ip\cdot x}\langle \Psi\,\pi^a(\hbox{\bf k})|T\{\eta(x)\,
\overline{\eta}(0)\}|\Psi\,\pi^b(\hbox{\bf k}')\rangle
\simeq-{1\over f_\pi^2}\delta^{ab}
{1\over 2}\Bigl(\Pi(p)+\gamma_5\Pi(p)\gamma_5\Bigr).
\label{soft2}
\end{equation}

The full expression for the leading correction to the correlator can be 
obtained by adding up (\ref{piinout},\ref{twopiin},\ref{twopiout}), using the
soft-pion approximation (\ref{soft2}). Making use of the expansion of the
pion field,
\begin{equation}
\pi^b(\hbox{\bf r})={1\over(2\pi)^{3/2}}\int {d^3\hbox{\bf k}
\over 2\omega_k}\left(a^b(\hbox{\bf k})\exp(i\hbox{\bf k}\cdot\hbox{\bf r})
+a^{b\dagger}(\hbox{\bf k})\exp(-i\hbox{\bf k}\cdot\hbox{\bf r})\right),
\end{equation}
the sum of these contributions can be conveniently expressed in the form
\begin{equation}
(2\pi)^3\langle\Psi|:{\textstyle{1\over 2}}\hbox{\boldmath$\pi$}
(0)^2:|\Psi\rangle\left(-{1\over f_\pi^2}\right)
{1\over 2}\Bigl(\Pi(p)+\gamma_5\Pi(p)\gamma_5\Bigr).
\end{equation}
Comparing this with the expression for the leading nonanalytic term in the 
sigma commutator (\ref{signonan}) above, we see that to first order in the 
density the pionic matrix element is
\begin{eqnarray}
(2\pi)^3\langle\Psi|:{\textstyle{1\over 2}}\hbox{\boldmath$\pi$}
(0)^2:|\Psi\rangle&\simeq&\rho(2\pi)^3\langle N|:{\textstyle{1\over 2}}
\hbox{\boldmath$\pi$}(0)^2:|N \rangle\nonumber\\
&\simeq&{\rho\sigma_{\pi{\scriptscriptstyle N}}^{(3)}\over m_\pi^2}.
\end{eqnarray}
Hence we may write the chiral expansion of the correlator (\ref{correl}) to
the order of interest, ${\cal O}(\rho m_\pi)$, as
\begin{equation}
\Pi(p)\simeq\overcirc\Pi(p)-{\xi\over 2}\Bigl(\overcirc\Pi(p)+\gamma_5 
\overcirc\Pi(p)\gamma_5\Bigr),
\label{correxp}
\end{equation}
where $\overcirc\Pi(p)$ denotes the correlator in the chiral limit.

The nucleon correlator in matter (\ref{correl}) can be decomposed into three
terms with diferent Dirac matrix structures\cite{nmsr},
\begin{equation}
\Pi(p)=\Pi^{(m)}(p)+\Pi^{(p)}(p)p\llap/+\Pi^{(u)}(p)u\llap/,
\end{equation}
where $u^\mu$ is a unit four-vector specifying the rest-frame of the matter.
From this we can see that only the piece $\Pi^{(m)}(p)$, corresponding to the
Dirac unit matrix, contributes to the ${\cal O}(\rho m_\pi)$ correction in
(\ref{correxp}). Hence to this order, the three pieces of the correlator are
\begin{eqnarray}
\Pi^{(m)}(p)&\simeq&(1-\xi)\overcirc \Pi^{(m)}(p),\nonumber\\
\Pi^{(p)}(p)&\simeq&\overcirc \Pi^{(p)}(p),\label{cpexp}\\
\Pi^{(u)}(p)&\simeq&\overcirc \Pi^{(u)}(p).\nonumber
\end{eqnarray}

The sum rules are obtained by equating the OPE and phenomenogical
representations of the correlator. To demonstrate the cancellation of the
leading chiral corrections to the nucleon energy, we need to examine the
chiral expansion of the pole and continuum pieces separately, as well as that
of the OPE. Consider first the OPE of the correlator. From (\ref{cpexp}) we
see that the leading correction to each condensate in the OPE of
$\Pi^{(m)}(p)$ takes the form of a factor of $1-\xi$, as for the quark
condensate in (\ref{qqlc}). This also follows from the fact that all these
condensates are matrix elements of operators that transform like $\overline
qq$ under chiral isospin rotations. In contrast the OPE's of $\Pi^{(p,u)}(p)$
have no corrections to this order, reflecting the fact that they involve
chirally invariant operators, such as $G_{\mu\nu}G^{\mu\nu}$ or $q^\dagger q$
respectively\cite{nmsr}. This common form for the leading corrections to all
the terms with a given Dirac structure was also noted in\cite{lccg} for
nucleon sum rules in vacuum.

By inserting a complete set of states with nucleon quantum numbers in the 
correlator $\overcirc \Pi(p)$ we can break (\ref{correxp}) up into 
chiral-limit nucleon-pole and continuum pieces:
\begin{equation}
\Pi(p)\simeq\left(1-{\xi\over 2}\right)\overcirc\Pi_{pole}(p)
-{\xi\over 2}\gamma_5 \overcirc\Pi_{pole}(p)\gamma_5
+\left(1-{\xi\over 2}\right)\overcirc\Pi_{cont}(p)
-{\xi\over 2}\gamma_5 \overcirc\Pi_{cont}(p)\gamma_5.
\label{corrpcexp}
\end{equation}
To interpret the various pieces of this expression in terms of the pole and
continuum parts of the correlator away from the chiral limit, we need to look
again at the double commutator in (\ref{soft}). Expanding this out gives four
terms
\begin{eqnarray}
\langle \Psi|\Bigl[Q_5^a,\Bigl[Q_5^b, T\{\eta(x)&&
\,\overline{\eta}(0)\}\Bigr]\Bigr]|\Psi)\rangle\label{dcomm}\\
=&&\langle \Psi|T\Bigl\{\left[Q_5^a,\left[Q_5^b,\eta(x)\right]\right]
\overline{\eta}(0)
+\eta(x)\Bigl[Q_5^a,\Bigl[Q_5^b,\overline{\eta}(0)\Bigr]\Bigr]
\nonumber\\
&&\qquad\quad+\Bigl[Q_5^a,\eta(x)\Bigr]\Bigl[Q_5^b,\overline{\eta}(0)
\Bigr]+\Bigl[Q_5^b,\eta(x)\Bigr]\Bigl[Q_5^a,\overline{\eta}(0)\Bigr]
\Bigr\}|\Psi\rangle.\nonumber
\end{eqnarray}

The first two terms contain double commutators of axial charges with the same
interpolating nucleon field. Applying the commutator (\ref{fchtr}) twice shows 
that these give rise to the correction term without factors of $\gamma_5$ in
(\ref{correxp}). If we insert a complete set of states into these terms 
expressed in the form (\ref{dcomm}) then we can see that they involve matrix
elements such as
\begin{equation}
\langle \Psi|\eta(x)|\Psi N(\hbox{\bf p})\rangle\langle\Psi N(\hbox{\bf p})|
\Bigl[Q_5^a,\Bigl[Q_5^b,\overline{\eta}(0)\Bigr]\Bigr]|\Psi\rangle,
\end{equation}
which comes from the soft-pion limit of
\begin{equation}
\langle \Psi|\eta(x)|\Psi N(\hbox{\bf p})\rangle\langle\Psi N(\hbox{\bf p})|
\overline{\eta}(0)|\Psi\,\pi^a(\hbox{\bf k})\pi^b(\hbox{\bf k}')\rangle.
\end{equation}
where $N(\hbox{\bf p})$ in these states denotes the probe nucleon created by
the interpolating fields. These terms are thus the soft-pion limits of the
pieces of (\ref{piinout},\ref{twopiin},\ref{twopiout}) where both pions couple
to the same interpolating field. They contribute to the residue of the nucleon
pole in (\ref{correl}) but do not alter its position, and hence they are
corrections in matter to the strength with which the interpolating field
couples to the nucleon. If we apply the same procedure that we used to get from
(\ref{piinout},\ref{twopiin},\ref{twopiout}) to (\ref{correxp}) to the pole
terms alone, we obtain
\begin{equation}
\Pi_{pole}(p)\simeq\left(1-{\xi\over 2}\right)\overcirc\Pi(p)_{pole}.
\end{equation}
Note that these terms provide the correction to the pole term in 
(\ref{correxp}) without factors of $\gamma_5$.

The second chiral-limit pole term in (\ref{correxp}), with $\gamma_5$'s, is
in fact part of the continuum of the correlator away from the chiral limit.
Like the similar term in the expression for the vacuum correlator\cite{lccg},
this arises from the last two terms in (\ref{dcomm}), which are the soft-pion
limits of terms such as
\begin{equation}
\langle \Psi|\eta(x)|\Psi N(\hbox{\bf p})\pi^b(\hbox{\bf k}')\rangle
\langle\Psi N(\hbox{\bf p})\pi^a(\hbox{\bf k})|\overline{\eta}(0)
|\Psi\rangle,
\end{equation}
where one pion couples to each interpolating field. Their form shows that they
arise from the lowest-energy part of the $\pi N$ continuum where the pion
scatters from one of the nucleons in the matter. 

Hence to the order of interest, the phenomenological representation of the
correlator can be written
\begin{equation}
\Pi(p)\simeq\Pi_{pole}(p)-{\xi\over 2}\gamma_5\Pi_{pole}(p)\gamma_5
+\left(1-{\xi\over 2}\right)\overcirc\Pi_{cont}(p)
-{\xi\over 2}\gamma_5 \overcirc\Pi_{cont}(p)\gamma_5,
\label{corrphen}
\end{equation}
where the last three terms form the continuum piece. For a nucleon in the
presence of mean scalar and vector potentials, the pole term of the correlator
in matter has the form\cite{nmsr}
\begin{equation}
\Pi_{pole}(p)=-\lambda^{*2}{p\llap/+M^*+V\gamma_0\over 2E(\hbox{\bf p})
[p^0-E(\hbox{\bf p})]},
\label{pole}
\end{equation}
where we have chosen to work in the rest frame of the matter. In (\ref{pole})
$M^*$ is the nucleon mass modified by the scalar potential, $V$ is the vector
potential and
\begin{equation}
E(\hbox{\bf p})=\sqrt{\hbox{\bf p}^2+M^{*2}}+V.
\end{equation}
Here $\lambda^*$ denotes the strength with which the interpolating field
couples to the nucleon in matter.

The sum rules are obtained by equating the pieces of (\ref{corrphen}) with the
corresponding ones of the OPE and then integrating over $p$ with some
convenient weighting function, such as the Borel transform\cite{svz,nmsr}.
For the nucleon in matter, this gives three sum rules corresponding to the 
three Dirac structures:
\begin{eqnarray}
-\left(1-{\xi\over2}\right)\lambda^{*2}M^*\int d^4p\,{w(p)\over 
2E(\hbox{\bf p})[p^0-E(\hbox{\bf p})]}&\simeq&(1-\xi)\int d^4p\, w(p)\left[
\overcirc\Pi^{(m)}_{OPE}(p)-\overcirc\Pi^{(m)}_{cont}(p)\right],\nonumber\\
-\left(1+{\xi\over2}\right)\lambda^{*2}\int d^4p\,{w(p)\over 2E(\hbox{\bf p})
[p^0-E(\hbox{\bf p})]}&\simeq&\int d^4p\, w(p)\left[\overcirc\Pi^{(p)}_{OPE}(p)
-\overcirc\Pi^{(p)}_{cont}(p)\right],\label{sumrules}\\
-\left(1+{\xi\over2}\right)\lambda^{*2}V\int d^4p\,{w(p)\over 2E(\hbox{\bf p})
[p^0-E(\hbox{\bf p})]}&\simeq&\int d^4p\, w(p)\left[\overcirc\Pi^{(u)}_{OPE}(p)
-\overcirc\Pi^{(u)}_{cont}(p)\right].\nonumber
\end{eqnarray}
By taking ratios of these sum rules one can see that to order $\rho m_\pi$
there are no chiral corrections to either $M^*$ or $V$. The self-energy of a
nucleon in matter thus contains no term of this order, in agreement with the
chiral counting rules\cite{wein,birqq}.

We note in passing that the form of the cancellation between the pieces of
this order in the condensates on the OPE side of the sum rules with pieces
from the phenomenological side is the same as that for the leading chiral
corrections to the sum rule in vacuum\cite{lccg} and the order $T^2$ pieces at
finite temperature\cite{el,koi}. This reflects the fact that all of these
contributions are produced by low-momentum pions, whose interactions with the
probe nucleon are subject to the same constraints due to chiral symmetry.

The leading nonanalytic term in the sigma commutator
$\sigma_{\pi{\scriptscriptstyle N}}^{(3)}$ is about $-25$ MeV, while the full
value is $\sigma_{\pi{\scriptscriptstyle N}} =45\pm 7$ MeV\cite{gls}. The
effects discussed here are thus significant on the scale of the change in the
quark condensate in matter given by (\ref{ratio}). In particular, the factor
of $1-\xi$ in the unit-matrix piece of the OPE, as in (\ref{cpexp}), is thus
$\sim 1.2$. If one keeps the terms of this order on the phenomenological side
that cancel this out, as in (\ref{sumrules}), then one would expect to see a
significant additional decrease in the nucleon mass, on top of that already
found in\cite{nmsr}. However one should remember that the size of the decrease
in the mass is also sensitive to assumptions about other condensates, in
particular the four-quark condensate\cite{nmsr}.

Finally, our results show that the factorisation ansatz often used for
the four-quark condensate cannot be valid in matter. That condensate, usually
denoted $\langle\Psi|(\overline qq)^2|\Psi\rangle$, appears in the OPE of
$\Pi^{(p)}(p)$ and so is a matrix element of a chirally invariant operator.
Replacing it by the factorised expression 
$\langle\Psi|\overline qq|\Psi\rangle^2$, which is not chirally invariant, 
introduces a term of order $\rho m_\pi$. Such a term is inconsistent with the
chiral expansion (\ref{cpexp}) of the correlator and cannot be cancelled by
pionic contributions from the phenomenological side of the correlator in
the way that we have seen happens for the ${\cal O}(\rho m_\pi)$ term in
the two-quark condensate.

\section*{Acknowledgements}

We are grateful to J. A. McGovern for a critical reading of the manuscript.
This work was supported by the EPSRC and PPARC.


\begin{thebibliography}{999}
\bibitem{csrevs}C. Adami and G. E. Brown, Phys. Reports {\bf 224} (1993) 1;\\
M. C. Birse, J. Phys.\ G: Nucl.\ Part.\ Phys.\ {\bf 20} (1994) 1537;\\
G. E. Brown and M. Rho, hep-ph/9504250, to be published in Phys.\ Reports.
\bibitem{svz}M. A. Shifman, A. I. Vainshtein and V. I. Zakharov, 
Nucl.\ Phys.\ {\bf B147} (1979) 385, 448, 519.
\bibitem{nmsr}T. D. Cohen, R. J. Furnstahl and D. K. Griegel, Phys.\ Rev.\ 
Lett.\ {\bf 67} (1991) 961;\\
T. D. Cohen, D. K. Griegel and R. J. Furnstahl, Phys.\ Rev.\ {\bf C46} (1992) 
1507;\\
X. Jin, T. D. Cohen, R. J. Furnstahl and D. K. Griegel, Phys.\ Rev.\ {\bf C47}
(1993) 2882;\\
X. Jin, M. Nielsen, T. D. Cohen, R. J. Furnstahl and D. K. Griegel, 
Phys.\ Rev.\ {\bf C49} (1995) 364.
\bibitem{dirph}B. D. Serot and J. D. Walecka, Adv.\ Nucl.\ Phys.\ {\bf 16} 
(1986) 1;\\
S. J. Wallace, Annu.\ Rev.\ Nucl.\ Part.\ Sci.\ {\bf 37} (1987) 267.
\bibitem{dl}E. G. Drukarev and E. M. Levin, Nucl.\ Phys.\ {\bf A511} (1990)
679; {\bf A516} (1990) 715(E). 
\bibitem{cfg}T. D. Cohen, R. J. Furnstahl and D. K. Griegel, Phys.\ Rev.\ {\bf
C45} (1992) 1881. 
\bibitem{gl}J. Gasser, Ann.\ Phys.\ {\bf 136} (1981) 62;\\
J. Gasser and H. Leutwyler, Phys.\ Reports {\bf 87} (1982) 77.
\bibitem{wein}S. Weinberg, Phys.\ Lett.\ {\bf 251} (1990) 288;
Nucl.\ Phys.\ {\bf B363} (1991) 1; Phys.\ Lett.\ {\bf B295} (1992) 114.
\bibitem{orvk}C. Ordo\~nez, L. Ray and U. van Kolck, Phys.\ Rev.\ Lett.\ {\bf
72} (1994) 1982; hep-ph/9511380.
\bibitem{birqq}M. C. Birse, University of Manchester preprint MC/TH 96/09, 
hep-ph/9602266.
\bibitem{el}V. L. Eletsky, Phys.\ Lett.\ {\bf B245} (1990) 229.
\bibitem{koi}Y. Koike, Phys.\ Rev.\ {\bf D48} (1993) 2313.
\bibitem{lccg}S. H. Lee, S. Choe, T. D. Cohen and D. K. Griegel, 
Phys.\ Lett.\ {\bf B348} (1995) 263.
\bibitem{bmcg}M. C. Birse and J. A. McGovern, Phys.\ Lett.\ {\bf B292} (1992) 
242;\\
I. Jameson, A. W. Thomas and G. Chanfray, J. Phys.\ G: 
Nucl.\ Part.\ Phys.\ {\bf 18} (1992) L159.
\bibitem{iof}B. L. Ioffe, Nucl.\ Phys.\ {\bf B188} (1981) 317; {\bf B191} 
591(E).
\bibitem{dgh}J. F. Donoghue, E. Golowich and B. R. Holstein, {\it Dynamics
of the standard model} (Cambridge University Press, Cambridge, 1992).
\bibitem{dei}M. Dey, V. L. Eletsky and B. L. Ioffe, Phys.\ Lett.\ {\bf B252} 
(1990) 620.
\bibitem{gls}J. Gasser, H. Leutwyler and M. E. Sainio, Phys.\ Lett.\ {\bf B253}
(1991) 252, 260.

\end{thebibliography}
\end{document}